\documentclass[aps,pra,reprint,longbibliography,floatfix]{revtex4-2}

\usepackage{graphicx}
\usepackage{amsmath}
\usepackage{amssymb}
\usepackage{bbm}
\usepackage{braket}
\usepackage{quantikz}
\usepackage{silence}
\usepackage[normalem]{ulem}
\usepackage{hyperref}

\hypersetup{colorlinks=true,linkcolor=blue,citecolor=blue,urlcolor=blue}

\begin{document}

\title{
Experimentally Extending Quantum Kernel Learning to Quantum Data by NMR
}

\author{Vivek Sabarad}
\email{vivek.sabarad@students.iiserpune.ac.in}
\author{Vishal Varma}
\email{vishal.varma@students.iiserpune.ac.in}
\author{T. S. Mahesh}
\email{mahesh.ts@iiserpune.ac.in}
\affiliation{Department of Physics and NMR Research Center,\
Indian Institute of Science Education and Research, Pune 411008, India}

\begin{abstract}
Quantum kernel learning (QKL) promises efficient machine learning by encoding feature maps onto exponentially large Hilbert spaces inherent in quantum systems. Using the liquid-state nuclear magnetic resonance (NMR) platform, we implement and benchmark QKL for one-dimensional regression and two-dimensional classification tasks. We then classify entangling and non-entangling operators by extending QKL to handle parametrized or non-parameterized operator inputs. We first compute the kernel numerically for a double-layered star system and then experimentally validate it on a 3-qubit NMR register. QKL provides a practical route to compare operators on native quantum hardware without expensive tomography protocols. Our results confirm the superiority of QKL over other classical methods for processing quantum data, thereby highlighting its ability to capture the inherent structure of quantum space and to extend its domain of operation beyond the training domain by exploiting symmetries in the operator space.
\end{abstract}

\maketitle

\section{Introduction}
Quantum machine learning (QML) lies at the intersection of quantum information science and statistical learning, promising to harness the structure of quantum systems for processing and learning from data \cite{Biamonte2017, Schuld2015, Jerbi2024} with wide-ranging applications in quantum many-body physics \cite{Cho2024, Lorente2022, Wu2023quantumphase, Uvarov2020}, quantum information processing \cite{Bisio2010, Peters2021, Huang2025}, quantum optics and imaging \cite{Xiao2023}, mathematical physics \cite{Paine2023}, high-energy physics \cite{nagano2023, Wu2021, Fadol2024} and many real-world applications \cite{Krunic2022, Miroszewski2024, Silva2024, Miroszewski2023}. Broadly, QML approaches fall into four paradigms \cite{Dunjko2016, Pira2023}: quantum-inspired classical algorithms for classical data (CC), quantum algorithms for classical data (QC), classical algorithms for quantum data (CQ), and quantum algorithms for quantum data (QQ). While early successes focused on the first three, recent attention has shifted toward using quantum systems to learn from quantum data. Theoretical research predicted that learning from such data can yield exponential benefits \cite{huang2022, Caro2023}, while numerical and experimental studies hinted at superior generalizability \cite{nagano2023, Bisio2010, Beer2020, Gibbs2024, Uvarov2020, Caro2022, Wu2023quantumphase}, i.e., the ability to predict beyond the training domain. 

While classical kernel methods underpin many reliable classical machine learning models \cite{scholkopf2002learning, shawe2004kernel}, the key challenge is in creating expressive yet trainable models from limited quantum data. Quantum kernel learning (QKL) methods offer a solution by mapping data onto high-dimensional Hilbert spaces via quantum circuits, allowing linear algorithms to identify nonlinear patterns with guaranteed generalizability and trainability \cite{schuld2019naturalqkernel}.
Foundational quantum algorithms for linear systems, principal component analysis, and support vector machines illustrate this potential \cite{Lloyd2009, Lloyd2014, Rebentrost2014}. Recent advances in variational quantum circuits and QKL have sharpened the theoretical picture \cite{Cerezo2022, Mangini2021, cerezo2021variational, Tacchino2021, Banchi2021, jerbi2023beyond, huang2021power}, while experimental demonstrations on superconducting hardware have confirmed the feasibility and potential of quantum feature spaces \cite{havlicek2019supervised}. Within this framework, QKL methods have emerged as a class of models that can offer practical quantum advantage \cite{schuld2019naturalqkernel, schuld2021quantum, Thanasilp2024, jager2023, Gentinetta2024}, with various use cases \cite{nagano2023, Wu2023quantumphase, Paine2023, Peters2021, Lorente2022, Wu2021, Caro2023, Krunic2022, Miroszewski2024, Silva2024, Miroszewski2023}.

The most compelling use case for QML involves \textit{quantum data} and has been primarily implemented via quantum neural networks \cite{nagano2023, Beer2020, Gibbs2024, Uvarov2020}. 
However, experimental progress, until now, has focused mainly on \textit{classical data} realized in NMR \cite{Li2015, kusumoto2021}, photonic \cite{bartkiewicz2020}, and neutral-atom \cite{Albrecht2023} architectures. 
For example, the solid-state NMR QKL of Kusumoto \textit{et al.} \cite{kusumoto2021} was limited to learning with classical data (QC).  
While it is well known that QKL can be much simpler to train than quantum neural networks in terms of resources \cite{schuld2021quantum}, an experimental demonstration of the former operating directly on quantum inputs has remained elusive to date.

Here, we report an experimental realization of QKL on a liquid-state NMR processor with star-topology registers \cite{jones2009magnetic,doi:10.7566/JPSJ.89.054001, Mahesh2021StartopologyRN}. 
While the solid-state NMR kernel \cite{Baum1985} employed by Kusumoto \textit{et al.} \cite{kusumoto2021} offers large clusters of spins correlated via an average double-quantum Hamiltonian of a multi-pulse RF sequence, the liquid-state NMR system, in principle, offers precisely controllable expressivity of kernels via gate-based quantum control over a register of definite size. More importantly, it allows us to devise an implementation protocol that can also handle quantum data, thus realizing the QQ paradigm of QML. 

In Sec. \ref{sec:theory}, we introduce and describe the theory behind classical and quantum kernels. In Sec. \ref{sec:QC}, we demonstrate the QC paradigm by realizing QKL on one-dimensional regression and two-dimensional classification problems using classical data. 
In Sec. \ref{sec:QQ}, we realize the QQ paradigm by extending QKL to quantum data. Here, we feed the kernel with unitary operators as input and show that it can classify entangling and non-entangling operators with high accuracy, generalizing cleanly to unseen data by capturing the symmetry structure of the quantum operator space. 
And finally, we close with discussions and conclusions in Sec. \ref{sec:discussion_conclusion}.

\section{Theory} \label{sec:theory}
\subsection{Kernel Methods \label{ssec:kernel_methods}}
{
Consider a dataset $\{ (\mathbf{x}_i, y_i) \}_{i=1}^N$ of $N$ observations, where $\mathbf{x}_i \in \mathbb{R}^d$ is the feature vector associated with the $i$-th observation and $y_i \in \mathbb{R}$ is the corresponding target value. A central problem in supervised machine learning (ML) is to infer the functional relationship between the inputs $\mathbf{x}_i$ and the targets $y_i$, commonly modeled as
\begin{equation}
    y_i = f(\mathbf{x}_i) + \epsilon_i,
    \label{eq:regression_model}
\end{equation}
where $f: \mathbb{R}^d \to \mathbb{R}$ is an unknown target function and $\epsilon_i$ accounts for observation noise. 
The simplest model for $f$ is a linear function of the input,
\begin{equation}
    f(\mathbf{x}_i) = w_0 + \sum_{k=1}^d w_k \, x_i^{(k)} = \mathbf{w}^\top \tilde{\mathbf{x}}_i,
    \label{eq:linear_model}
\end{equation}
with $\tilde{\mathbf{x}}_i = (1, x_i^{(1)}, \dots, x_i^{(d)})^\top$ and the noise $\epsilon_i$ has been absorbed by the bias $w_0$. While analytically convenient, linear models of this form cannot capture nonlinear dependencies between $\mathbf{x}_i$ and $y_i$, limiting their applicability to real-world data.
}

{
A natural strategy for extending the linear model of Eq.~\eqref{eq:linear_model} to the nonlinear setting is to first embed the input data into a higher- or even  infinite-dimensional \textit{feature space} $\mathcal{F}$ via a nonlinear feature map
\begin{equation}
    \phi: \mathbb{R}^d \to \mathcal{F}, \qquad \mathbf{x} \mapsto \phi(\mathbf{x}),
\end{equation}
and to subsequently fit a linear model \textit{in} $\mathcal{F}$,
\begin{equation}
    f(\mathbf{x}_i) = \mathbf{w}_\phi^\top \phi(\mathbf{x}_i).
    \label{eq:feature_space_model}
\end{equation}
Because $f$ remains linear in $\mathbf{w}_\phi$, the resulting model retains the convexity and analytical tractability of linear regression \cite{Book_Bishop_Pattern}.
The parameters $\mathbf{w}_\phi$ are obtained by minimizing the regularized empirical risk
\begin{equation}
    J(\mathbf{w}_\phi) = \sum_{i=1}^N \left\{y_i - \mathbf{w}_\phi^\top \phi(\mathbf{x}_i) \right\}^2 + \lambda  \mathbf{w}_\phi^\top \mathbf{w}_\phi,
    \label{eq:primal_objective}
\end{equation}
where $\lambda > 0$ is a regularization parameter that controls the trade-off between fitting the training data and controlling the norm of $\mathbf{w}_\phi$, thereby mitigating overfitting. Setting the gradient of Eq. \eqref{eq:primal_objective} with respect to $\mathbf{w}_\phi$ to zero,
\begin{equation}
    \frac{\partial J(\mathbf{w}_\phi)}{\partial \mathbf{w}_\phi}
    = -2 \sum_{i=1}^N \left\{ y_i - \mathbf{w}_\phi^\top \phi(\mathbf{x}_i) \right\} \phi(\mathbf{x}_i) + 2\lambda \mathbf{w}_\phi = \mathbf{0},
\end{equation}
shows that the optimal solution admits an expansion in terms of the training points,
\begin{equation}
    \!\!\!
    \mathbf{w}_\phi \!=\! \sum_{i=1}^N \alpha_i \phi(\mathbf{x}_i) = \mathbf{\Phi}^\top \boldsymbol{\alpha},
    \quad
    \alpha_i \!=\! \frac{1}{\lambda} \! \left\{ y_i - \mathbf{w}_\phi^\top \phi(\mathbf{x}_i) \right\},
    \label{eq:representer}
\end{equation}
{
where $\mathbf{\Phi}$ is the \textit{design matrix} with $i$-th row 
$\phi(\mathbf{x}_i)^\top$, 
and $\boldsymbol{\alpha} \in \mathbb{R}^N$ 
are the \textit{dual coefficients}, weights assigned to the training points \cite{Book_Bishop_Pattern}. This is a special case of a more general fact described by the \textit{representer theorem} \cite{Scholkopf2001}. For this type of regularized fitting problem, the best-fitting function can always be written as a weighted combination of the training points as in Eq.~\eqref{eq:representer}, no matter how large or complicated the feature space $\mathcal{F}$ is, even if it's infinite-dimensional.
}
}

\begin{figure}
    \centering
\includegraphics[width=\linewidth]{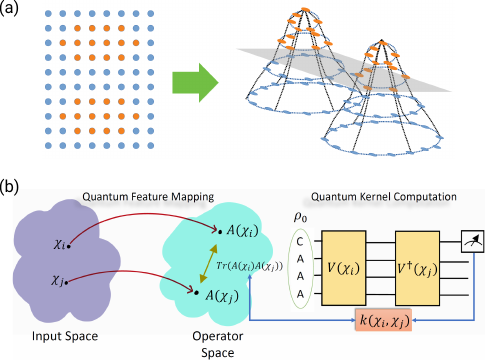}\\
    \caption{
    {
    (a) Illustrating a dataset requiring a nonlinear function to classify (left) and a feature space mapping onto a higher-dimensional space where it gets classified by a hyperplane.
    (b) Quantum feature mapping and computation protocol for QKL.
    }
    }
    \label{fig:abstract_kernel}
\end{figure}
{
Substituting Eq. \eqref{eq:representer} into Eq. \eqref{eq:primal_objective} recasts the optimization problem in terms of $\boldsymbol{\alpha}$ rather than $\mathbf{w}_\phi$,
\begin{align}
    J(\boldsymbol{\alpha}) =\;
    & \boldsymbol{\alpha}^\top \mathbf{\Phi}\mathbf{\Phi}^\top \mathbf{\Phi}\mathbf{\Phi}^\top \boldsymbol{\alpha}
    - 2\boldsymbol{\alpha}^\top \mathbf{\Phi}\mathbf{\Phi}^\top \mathbf{y}
    + \mathbf{y}^\top \mathbf{y} \nonumber \\
    & + \lambda \, \boldsymbol{\alpha}^\top \mathbf{\Phi}\mathbf{\Phi}^\top \boldsymbol{\alpha},
    \label{eq:dual_objective_raw}
\end{align}
where \(
\mathbf{y} = \left(y_1, y_2, \dots, y_N \right)^\top
\) is the vector of target values in the training dataset.
}
{
Crucially, $\mathbf{\Phi}$ and the feature map $\phi$ enter Eq.~\eqref{eq:dual_objective_raw} \textit{exclusively} through the inner products $\phi(\mathbf{x}_i)^\top \phi(\mathbf{x}_j)$, and never in isolation. This observation motivates the definition of the \textit{kernel function}
\begin{equation}
    k(\mathbf{x}_i, \mathbf{x}_j) := \phi(\mathbf{x}_i)^\top \phi(\mathbf{x}_j),
    \label{eq:kernel_def}
\end{equation}
and the associated \textit{kernel} (or \textit{Gram}) \textit{matrix} $\mathbf{K} \in \mathbb{R}^{N \times N}$, $K_{ij} = k(\mathbf{x}_i, \mathbf{x}_j)$, so that $\mathbf{K} = \mathbf{\Phi}\mathbf{\Phi}^\top$. In terms of $\mathbf{K}$, Eq.~\eqref{eq:dual_objective_raw} simplifies to
\begin{equation}
    J(\boldsymbol{\alpha}) = \boldsymbol{\alpha}^\top \mathbf{K}^2 \boldsymbol{\alpha} - 2\boldsymbol{\alpha}^\top \mathbf{K}\mathbf{y} + \mathbf{y}^\top\mathbf{y} + \lambda\, \boldsymbol{\alpha}^\top \mathbf{K} \boldsymbol{\alpha},
\end{equation}
whose minimizer with respect to $\boldsymbol{\alpha}$ satisfies the normal equations
\begin{equation}
    \boldsymbol{\alpha} = \left( \mathbf{K} + \lambda \mathbf{I}_N \right)^{-1} \mathbf{y},
    \label{eq:dual_solution}
\end{equation}
where $\mathbf{I}_N$ is the $N$ dimensional identity matrix \cite{Book_Bishop_Pattern}.
Given $\boldsymbol{\alpha}$, predictions for a new input $\mathbf{x}$ follow directly from Eq.~\eqref{eq:representer} and \eqref{eq:kernel_def} as
\begin{equation}
    f(\mathbf{x}) = \sum_{i=1}^N \alpha_i \, \phi(\mathbf{x}_i)^\top \phi(\mathbf{x}) = \sum_{i=1}^N \alpha_i \, k(\mathbf{x}_i, \mathbf{x}).
    \label{eq:prediction}
\end{equation}
}

{
Eqs. (\ref{eq:dual_solution}--\ref{eq:prediction}) show that the entire learning problem, both training and prediction, can be expressed solely in terms of the kernel function $k(\cdot, \cdot)$, without ever constructing or evaluating $\phi(\mathbf{x})$ explicitly. This substitution of inner products by a directly computable kernel function is known as the \textit{kernel trick} \cite{scholkopf2002learning}. It is applicable whenever the underlying algorithm can be formulated exclusively in terms of inner products between data points, a property shared by numerous ML methods \cite{hofmann2008kernel}.
}

{
For classification, the kernel is used to compute the decision boundary between different classes. The support vector machine (SVM) classifier \cite{cortes1995} uses the kernel matrix to find the optimal hyperplane that separates the data points of different classes in the feature space $\mathcal{F}$ (see Fig.  \ref{fig:abstract_kernel} (a)). The predictions for new input data points involve computing kernel values over all training data points and applying the learned coefficients.}

{Classical kernels, however, require the input to be a finite-dimensional vector of real numbers. For classical data, this limits the feature space to what a classical computer can efficiently compute. For quantum data such as unitary operators, either a parameterization or tomography becomes necessary before feeding it into a classical kernel. The former loses information about the operator structure, and the latter scales exponentially with system size. QKL not only exploits the exponentially scaling Hilbert space but also compares operators directly, as discussed in the following section.}

\subsection{Quantum feature maps and quantum kernels \label{sec:qfeaturemaps}}
QKL extends classical kernel techniques by embedding input data into the space of operators acting on a quantum system, thereby exploiting the rich algebraic structure of quantum mechanics without requiring a fault-tolerant, large-scale quantum computer \cite{schuld2021quantum}. 
QKL comprises two steps: a quantum feature mapping, which encodes each data point into an operator on the Hilbert space of the quantum system, and a kernel computation that compares two input operators \cite{Wang2024}. Whereas most existing QKL constructions are restricted to encoding classical data within the ambit of the QC paradigm, here we consider the less-explored QQ paradigm, in which the input data can be a quantum operator. This generalization is possible because, as shown below, both the feature mapping and the kernel evaluation are carried out natively on the quantum processor without requiring a classical representation of the input data.

For each input $\chi_i$, we define the associated feature-space operator
\begin{equation}
    A(\chi_i) = V(\chi_i)\, A_0 \, V^\dagger(\chi_i),
    \label{eq:QFM}
\end{equation}
where $A_0$ is a fixed reference operator (observable) on the system, and $V(\chi_i)$ is a unitary encoding the data point $\chi_i$. 
The design of $V(\chi_i)$ is subject to an important trade-off. While a more expressive encoding unitary increases the effective dimension of the feature space and hence the representational power of the model, overly expressive feature maps have been shown to cause the kernel values $k(\chi_i,\chi_j)$ to concentrate exponentially (in the number of qubits) around a fixed value, independent of $(\chi_i, \chi_j)$ \cite{Thanasilp2024}. In the concentrated regime, the kernel becomes uninformative, and the associated model loses its ability to generalize, analogous to the barren-plateau phenomenon in variational quantum circuits. The choice of $V(\chi_i)$ must therefore balance expressivity against this concentration effect.

Given the operator-valued feature map of Eq.~\eqref{eq:QFM}, the natural analogue of the classical inner product $\phi(\mathbf{x}_i)^\top\phi(\mathbf{x}_j)$ of Eq.~\eqref{eq:kernel_def} is the Hilbert--Schmidt (Frobenius) inner product between operators,
\begin{equation}
    k(\chi_i, \chi_j) = \mathrm{Tr}\left[ A(\chi_i)\, A(\chi_j) \right],
    \label{eq:kfrob}
\end{equation}
which reduces to $\mathrm{Tr}[A^\dagger B] $ for the Hermitian operators considered here, and is symmetric, $k(\chi_i,\chi_j) = k(\chi_j,\chi_i)$, and positive semi-definite by construction, so that Eq.~\eqref{eq:kfrob} defines a valid kernel 
\cite{Book_Bishop_Pattern}. 

Substituting Eq.~\eqref{eq:QFM} into Eq.~\eqref{eq:kfrob} and using the cyclic property of the trace, the kernel can be rewritten as
\begin{equation}
    k(\chi_i, \chi_j) = \mathrm{Tr}\left[ V^\dagger(\chi_j) V(\chi_i) \, A_0 \, V^\dagger(\chi_i) V(\chi_j) \, A_0 \right].
    \label{eq:kernel_cyclic}
\end{equation}

More generally, one may replace the bare reference operator $A_0$ in the state-preparation step by an initial state of the form $\rho_0 = B + a A_0$, where $a$ is a scalar and $B$ is any operator satisfying $[B, V(\chi_i)] = 0$ for all $\chi_i$, i.e.\ a background contribution left invariant by the encoding unitaries. In this work, we take $B = \mathbbm{1}/2^n$, the maximally mixed state on $n$ qubits, representing a uniform background population contributing only an additive constant to the kernel, which can be discarded.  Thus, up to a normalization constant, the kernel is expressed as,
\begin{equation}
    k(\chi_i, \chi_j) \propto \mathrm{Tr}\left[ V^\dagger(\chi_j) V(\chi_i) \, \rho_0 \, V^\dagger(\chi_i) V(\chi_j) \, A_0 \right].
    \label{eq:kernel_explicit}
\end{equation}
Eq.~\eqref{eq:kernel_explicit} admits a direct operational interpretation: the kernel is obtained by preparing the initial state $\rho_0$, applying the unitary $V(\chi_j)^\dagger V(\chi_i)$, and measuring the expectation value of the observable $A_0$,
without ever reconstructing $A(\chi_i)$ classically.
A schematic representation of the quantum feature mapping and kernel evaluation protocol is shown in Fig.~\ref{fig:abstract_kernel} (b).

\section{QC paradigm on a star-topology system
\label{sec:QC}}
\begin{figure*}[t]
    \centering
    \includegraphics[width=0.95\linewidth]{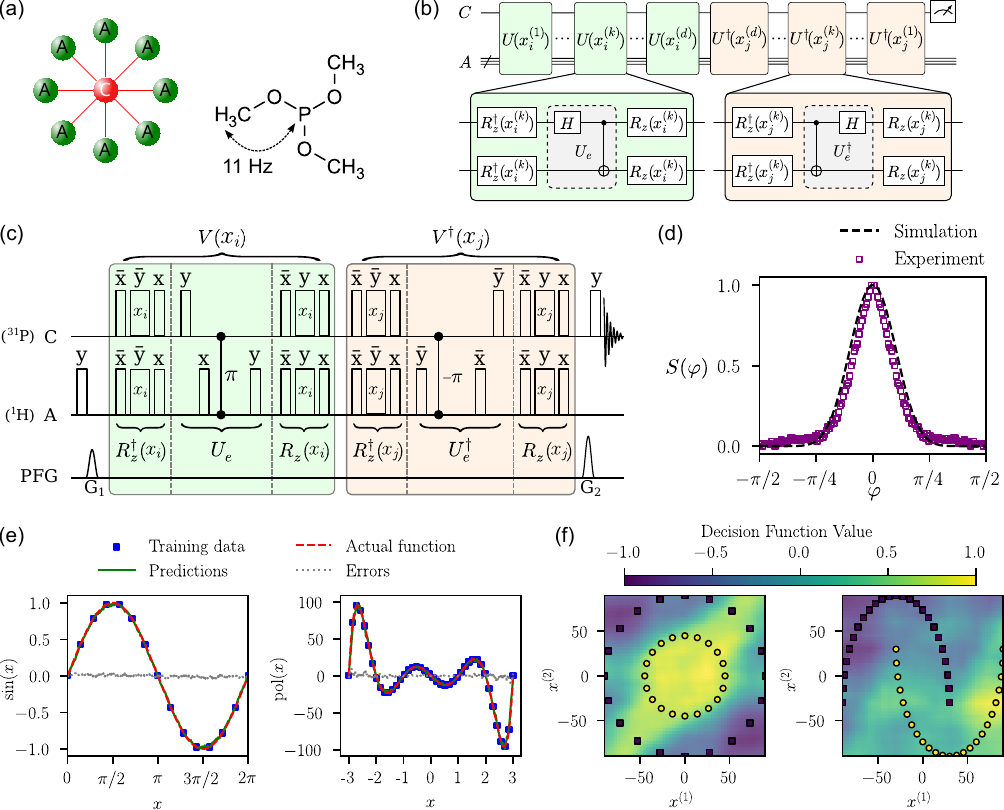}
    \caption{(a) Star-topology register and the molecular structure of trimethyl phosphite, a 10-qubit system with $J_\mathrm{HP}=11$ Hz, used for performing quantum feature mapping and QKL with classical data.  
    (b) Quantum circuit for encoding $d$-dimensional classical inputs $\mathbf{x}_i$ and $\mathbf{x}_j$ into unitaries $U(\mathbf{x}_i)$ and $U(\mathbf{x}_j)$, followed by measurement of $I_z^C$. 
    (c) {NMR pulse sequence to implement QKL with classical data. 
    Here, empty rectangles represent $90^\circ$ pulses with the phases mentioned on top.  The pulses corresponding to $x_{i(j)}$ rotations are indicated explicitly. 
    Vertical lines connecting two qubits represent $J$ evolutions by angles as indicated. 
    The pulse field gradients (PFGs) $G_1$ and $G_2$ are used to dephase coherences during initial state preparation and before the final detection pulse.
    }
    (d) Simulated and experimentally obtained kernels (see Eq. \eqref{eq:kernel_shifted}) for one-dimensional inputs plotted vs. $\varphi = x_i-x_j$. Assuming the symmetry of the kernel function, the left half is obtained by mirroring the experimentally extracted right half.
    (e) NMR QKL regression of the sine function with 15 training data points (left), and a 7th-degree polynomial function with 42 training data points (right). 
    (f) NMR QKL classification of two-dimensional tasks for the circles dataset (left) and the moons dataset (right). In both cases, the squares and circles represent training data points from two different classes. The background represents the decision functions learned by the SVM.
    }
    \label{fig:qc_all}
\end{figure*}

{
We start by benchmarking the QKL on classical input data using one-dimensional regression and two-dimensional classification tasks. 
We consider a star-topology register  with $n+1$ qubits consisting of a central qubit (C) coupled uniformly to $n$ magnetically equivalent ancillary qubits (A) via the Hamiltonian,
\begin{equation}
H_J = 2\pi J \sum_{i=1}^n I_z^CI_{z}^{A_i},
\end{equation}
where $J$ is the coupling constant and $I_z^{C(A_i)}$ is the $z$-component of the spin angular momentum for the central ($i$-th ancillary) qubit (see Fig. \ref{fig:qc_all} (a)).
Such a spin system naturally occurs in a variety of molecules and has been utilized for several spectroscopic as well as quantum information processing tasks \cite{Mahesh2021StartopologyRN}. 
}

\subsection{Feature map for encoding classical data}
To encode a $d$-dimensional input vector $\mathbf{x}_i = (x_i^{(1)}, x_i^{(2)}, \dots, x_i^{(d)}) \in \mathbb{R}^d$, we define the unitary
\begin{equation}
    V(\mathbf{x}_i) = \prod_{j=1}^d U(x_i^{(j)}) = \prod_{j=1}^d e^{-i x_i^{(j)} I_z} \, U_e \, e^{i x_i^{(j)} I_z},
    \label{eq:encoding_unitary}
\end{equation}
where $I_z = I_z^C + \sum_{i} I_z^{A_i}$ is the total $z$-angular-momentum operator, summed over all qubits in the register, and $U_e$ is a fixed entangling unitary. 
While $U_e$ can be realized in any register, the star-topology register allows highly efficient realization of $U_e$ with a Hadamard on C followed by a single CNOT operation between C and A (see the dashed box in Fig. \ref{fig:qc_all} (b)).

For a set of scalar inputs $\{x_i\}_{i=1}^N$ ($d=1$ in Eq. \eqref{eq:encoding_unitary}), 
\begin{equation}
    V(x_i) = e^{-i x_i I_z}\, U_e\, e^{i x_i I_z},
    \label{eq:1d_encoding_unitary}
\end{equation}
where $e^{\mp i x_i I_z}$ implements a collective rotation of all qubits about the $z$ axis. 
Choosing the reference operator $A_0 = I_z^C$, the general kernel of Eq. \eqref{eq:kfrob} can now be written as
\begin{equation}
    k(x_i, x_j) = \mathrm{Tr}\!\left[ V(x_i)\, I_z^C\, V^\dagger(x_i)\, V(x_j)\, I_z^C\, V^\dagger(x_j) \right].
    \label{eq:kernel}
\end{equation}
Substituting Eq.~\eqref{eq:1d_encoding_unitary} into Eq.~\eqref{eq:kernel} and using that $I_z^C$ commutes with the collective rotation $e^{\pm i x I_z}$, the two collective rotations combine into a single relative rotation, and cyclicity of the trace removes the remaining dependence on $x_i$ and $x_j$ individually. The kernel is therefore a function of the difference $\varphi \equiv x_i - x_j$ alone,
\begin{equation}
    k(x_i, x_j) = \mathrm{Tr} \left[U_e^\dagger e^{-i\varphi I_z} U_e I_z^C U_e^\dagger e^{i\varphi I_z} U_e I_z^C \right].
    \label{eq:knmr1d}
\end{equation}
By an appropriate choice of initial state (Sec.~\ref{sec:theory} B), the kernel of Eq.~\eqref{eq:knmr1d} becomes proportional to the experimentally accessible signal
\begin{equation}
    S(\varphi) = \mathrm{Tr} \left[U_e^\dagger e^{-i\varphi I_z} U_e \;\rho_0\; U_e^\dagger e^{i\varphi I_z} U_e I_z^C \right] \propto k(x_i,x_j).
    \label{eq:kernel_shifted}
\end{equation}

For $d>1$ inputs, the successive rotations in Eq.~\eqref{eq:encoding_unitary} no longer commute with a single collective phase, and the kernel loses the shift-invariance of Eq.~\eqref{eq:knmr1d}, becoming a genuine function of $\mathbf{x}_i$ and $\mathbf{x}_j$ separately (Appendix \ref{app:2dkernel}).

\subsection{NMR QKL with classical data}
\label{sec:cq_exp_implementation}
We implement the kernel computation on a star-topology register using the NMR spin system of trimethyl phosphite (see Fig. \ref{fig:qc_all} (a)), in which the $^{31}$P spin acts as the central qubit (C) and the nine magnetically equivalent $^1$H spins act as the ancillary qubits (A). All NMR experiments were performed on an 11.7 Tesla Bruker Avance-III NMR spectrometer with a QXI probe whose thermostat was set to 298 K. 
The longitudinal spin relaxation time constants ($T_1$)  for the $^{31}$P and $^1$H spins are 9.7 s and 5.6 s, respectively. 

The thermal equilibrium state of the $10$-spin STR is
\begin{equation}
    \rho_{\mathrm{th}} = \frac{1}{2^{10}}\mathbbm{1} + \epsilon_C I_z^C + \epsilon_A \sum_{i=1}^{9} I_z^{A_i},
    \label{eq:rho_eq}
\end{equation}
where $\epsilon_{C(A)}$ denotes the equilibrium polarizations of the central (ancillary) spins \cite{Mahesh2021StartopologyRN}.
We implement the quantum feature map (Eq. \eqref{eq:QFM}) for a given classical input $\mathbf{x}_i$ with the reference operator $A_0 = I_z^C$. 
The full pulse sequence for realizing the quantum feature map and computing the kernel is shown in Fig. \ref{fig:qc_all} (c).
In the following, we assess the performance of the NMR QKL on tasks such as regression and classification.

For one-dimensional classical inputs, we compute the kernel function at equally spaced intervals between 0 and $\pi/2$. 
From Fig. \ref{fig:qc_all} (d), we observe that the experimental kernel function (using Eq. \eqref{eq:kernel_shifted}) agrees very well with the simulated kernel function.  Using the experimental kernel with ridge regression, we perform QKL of the target function and predict the value for an unseen input.

The results of one-dimensional regression tasks are shown in Fig. \ref{fig:qc_all} (e).  
First, we show regression for a sine function with one period from 0 to $2\pi$.  With only 15 training points, we achieve an RMS accuracy of 99.1\%.
Then we show the regression 7th-degree polynomial $y = (x-3)(x-2)(x-1)\;x\;(x+1)(x+2)(x+3)$ with 42 training points, achieving an RMS accuracy of 98.8\%.

To implement QKL with two-dimensional classical data, we experimentally compute the kernels (see Eqs. \eqref{eq:kernel_explicit}, \eqref{eq:encoding_unitary}, and Appendix \ref{app:2dkernel}), and then perform two classification tasks: a Circles dataset and a Moons dataset \cite{scikit-learn-datasets}, as shown in Fig. \ref{fig:qc_all} (f). 
The hinge loss values \cite{cortes1995}, which quantify errors in these classifications, are 0.15 and 0.08, respectively. See Appendix \ref{app:workflow} for the detailed workflow involved in the above tasks. These results show that QKL is fairly reliable and can perform standard ML tasks, suggesting that we can extend the approach to quantum data.

\section{QQ paradigm: QKL directly from unitary operators}\label{sec:QQ}
\begin{figure*}
    \centering
    \includegraphics[width=\linewidth]{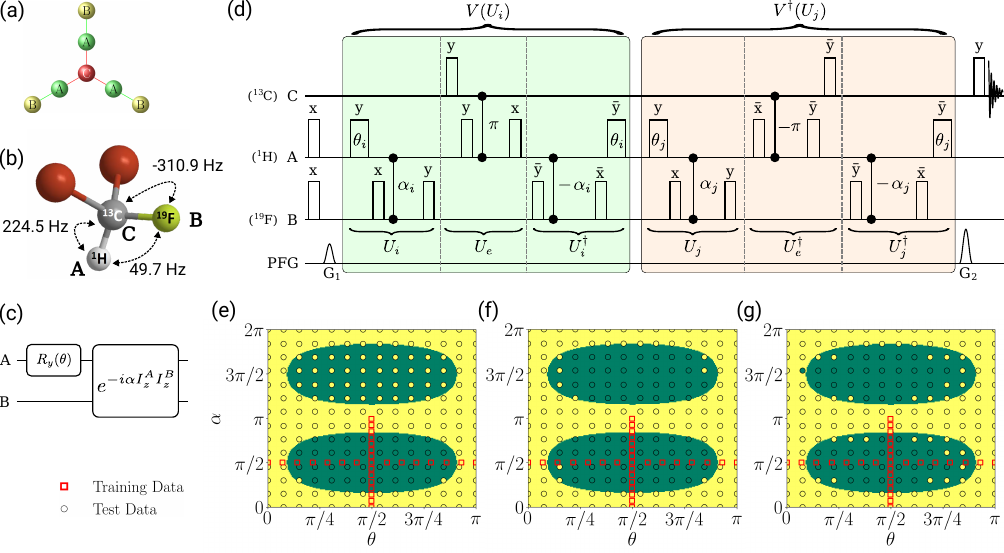}
    \caption{
    (a) The geometry of the double-layer star topology spin register used for the entanglement classification task. 
    (b) The molecular structure of the dibromofluoromethane used to realize one branch of the double-layer register. 
    (c) Quantum circuit for generating input unitaries $U_i$ parameterized by $\theta$ and $\alpha$.
    (d) {NMR QKL computation for classification of two-qubit unitaries. 
    The empty rectangles represent $90^\circ$ pulses with the phases mentioned on top.  The pulses corresponding to $\theta_{i(j)}$ rotations are indicated explicitly. 
    Vertical lines connecting two qubits represent $J$ evolutions by angles as indicated. 
    The pulse field gradients (PFGs) $G_1$ and $G_2$ are used to dephase the coherences during initial state preparation and before the final detection pulse.
    }
    (e) Numerically simulated classification using a classical Gaussian kernel trained on the parameter space $(\theta,\alpha)$ showing an accuracy less than 50\%.
    (f) Numerically simulated QKL classification of entangling vs. non-entangling unitaries, showing accuracy of {94\%}.  
    (g) Experimental NMR QKL classification showing an accuracy of {85\%}. 
    In (e-g), the squares and circles correspond to training and test data, respectively; the area inside and outside the elliptical patch corresponds to entangling and nonentangling unitaries applied to the thermal initial state $ \rho_{th}$. 
    }
    \label{fig:entanglement_classification}
\end{figure*}
We now turn to our main objective: demonstrating the QQ paradigm, 
in which both the feature map and the input data are genuinely quantum. As a representative task, we consider \textit{entanglement classification}, a binary classification problem in which a unitary operator $U_i \in SU(4)$, acting on a fixed two-qubit mixed initial state $\rho = \frac{1}{4}\mathbbm{1} + \epsilon (I_z^{A} +  I_z^{B})$, is to be classified as an entangling or non-entangling transformation.  Here, a unitary $U_i$ is labeled \textit{entangling} if logarithmic negativity of the resulting state $\rho_i = U_i \rho U_i^\dagger$ is strictly positive, and \textit{non-entangling} otherwise.
This is a nontrivial problem in general \cite{Serrano-Ensastiga2023}, which we address using the QKL.

\subsection{Feature map for encoding quantum data}
To encode and classify unitary-valued quantum inputs, we use a double-layered star-topology spin register as shown in Fig. \ref{fig:entanglement_classification} (a). 
For an arbitrary input unitary $U_i$, acting on the subspace formed by the A and B qubits of a given branch, we define the encoding unitary
\begin{equation}
    V_i = U_i^\dagger U_e U_i,
    \label{eq:qt_encoding_unitary}
\end{equation}
where $U_e$ is a fixed entangling unitary consisting of a Hadamard gate on the C qubit followed by a CNOT-like operation between the C qubit and each A qubit. 
{
This way of encoding the input unitaries ensures that any unitary $U_i \in SU(4)$ can be encoded onto the quantum processor, including those that do not have a simple closed-form parameter description. 
The kernel $k(U_i, U_j)$ defined via the Frobenius inner product of Eq.~\eqref{eq:kfrob} measures the overlap between $U_i$ and $U_j$, and is evaluated using the general expression of Eq. \eqref{eq:kernel_explicit}. 
We take the reference operator $A_0 = I_z^C$ and the initial state $\rho_0 \propto I_z^C$ as in the previous case.
With this encoding, the general kernel expression of Eq. \eqref{eq:kernel_explicit} becomes
\begin{align}
    k(U_i, U_j) \propto& \text{Tr}\!\left[ V_j^\dagger V_i I_z^C V_i^\dagger V_j I_z^C \right],
\end{align}
which is equal to the measured expectation value of $I_z^C$ after applying the transformation $U_j^\dagger U_e^\dagger U_j 
    U_i^\dagger U_e U_i$ on $I_z^C$.
}

\subsection{NMR QKL with quantum data \label{sec:qq_exp}}
We used 13C-labeled dibromofluoromethane dissolved in acetone-d6 to realize one branch of the double-layered star system (see Fig. \ref{fig:entanglement_classification} (a-b))\cite{Swathi2016}. 
The longitudinal relaxation time constants for $^{13}\mathrm{C}$, $^1\mathrm{H}$, and $^{19}\mathrm{F}$ are 1.9 s, 13.7 s, and 5.2 s, respectively.
The thermal state of the $3$-spin system is
\begin{equation}
    \rho_{\mathrm{th}} = \frac{1}{2^3}\mathbbm{1} +  \sum_{i=A,B,C} \epsilon_i I_{zi},
    \label{eq:rho_eq_dbfm}
\end{equation}
from which we prepare $\rho_0 \propto I_z^C$ by dephasing the magnetizations of $A$ and $B$ spins.
We generate synthetic data for training and testing by considering a parametrized quantum circuit as shown in Fig. \ref{fig:entanglement_classification} (c). 
The full pulse sequence for realizing the quantum feature map and computing the kernel is shown in Fig. 
\ref{fig:entanglement_classification} (d).
In the following, we discuss the performance of the NMR QKL for the classification of entangling vs. nonentangling unitaries.

We now compare the QQ paradigm with the CQ paradigm.
Fig. \ref{fig:entanglement_classification} (e) shows the results of numerical simulation of the CQ paradigm with a classical Gaussian kernel trained directly on the parameter space $(\theta,\alpha)$ \cite{scikit-learn}. The areas inside elliptical patches represent parameters $\left(\theta, \alpha\right)$ corresponding to entangling unitaries $U\left(\theta, \alpha\right)$, and those outside correspond to nonentangling unitaries. We train the model on 30 operators shown by squares, chosen exclusively from the lower half of the parameter space, and test the model on 196 operators sampled uniformly over the entire parameter space. 
While the classical kernel correctly identifies the entangling region local to the training data, i.e., in the lower patch, it fails entirely to identify the entangling region in the upper patch of the parameter space, resulting in the overall prediction accuracy of about 50\%. This illustrates a key limitation of the CQ paradigm when classical kernels are applied to quantum data.

We now switch to the QQ paradigm with the same sizes for the training and test data sets.
Fig. \ref{fig:entanglement_classification} (f) shows the results of numerical simulation for the $3$-branch system. 
Naively, one would again expect the prediction to hold only for the lower half of the parameter space.  However, the numerical simulation of the QKL classification yielded a prediction accuracy of 94\%.  In fact, the accuracy improved further with larger training sets and eventually saturated at about 99\%, as shown in the Appendix (see \ref{app:qq_accuracy_vs_tau}). 
It is interesting to note that although the model is trained only on the lower half of the parameter space, it also performs well on the upper half, well beyond the training region. 

Now we turn to the experimental implementation of QKL classification using the NMR register.  The results shown in Fig. \ref{fig:entanglement_classification} (g) indicate an accuracy of 85\% on the same size of datasets.  The experimental NMR QKL classification correctly identifies both the lower and upper entangling patches, thus demonstrating the successful realization of the QQ paradigm.
These results are consistent with the proposed advantages of learning directly from quantum data over classically parameterized operators \cite{Cerezo2022, huang2022, Caro2023}. First, we observe generalizability, i.e., a QKL model trained only on unitaries from the lower half of parameter space (Fig.~\ref{fig:entanglement_classification} (g)) successfully classifies unitaries from the untrained upper half, indicating that the kernel captures structure in the operator space beyond what is visible in the training distribution alone. Second, because unitary inputs are fed directly into the kernel model, it requires no finite classical parameterization of the input, underscoring the flexibility of working with quantum data natively.

\section{Discussions and  Conclusions \label{sec:discussion_conclusion}}
In this work, we experimentally implemented quantum kernel learning (QKL) methods on NMR-based spin registers for both classical and quantum data, demonstrating a practical route to access the high-dimensional feature spaces inherent in quantum systems. 
First, using a 10-qubit star-topology register, we encoded classical data into the $2^{10}$-dimensional Hilbert space via data-dependent unitary transformations and experimentally computed the resulting kernel, demonstrating that it supports standard machine learning tasks such as one-dimensional regression and two-dimensional classification. Extending the register to a double-layered star configuration, we further constructed a QKL method capable of classifying non-parameterized unitary operators directly as quantum inputs, and validated it through both numerical simulations and experiments on a 3-spin NMR register.
A well-trained QKL avoids resource-expensive quantum state tomography and is therefore promising for a wide range of classification tasks. 

Our results indicate that QKL is realizable on NISQ-era platforms and can process quantum data directly, without an explicit classical parameterization of inputs. This work thus opens new possibilities and marks a step toward more general quantum machine learning with potential near-term applications in quantum many-body physics, high-energy physics \cite{nagano2023}, and other emerging quantum technologies \cite{Peters2021, Lorente2022, Wu2023quantumphase}.

\section*{Acknowledgements}
The authors gratefully acknowledge discussions with Prof. Sreejith G. J., Arijit Chatterjee, and Keshav V. We also used ChatGPT to help generate simulation code for this study. V. V. acknowledges the University Grants Commission (UGC) fellowship MR22031373. T.S.M. acknowledges funding from I-HUB QTF.

\appendix

\section{Kernel Extraction for Two-dimensional Classical Data}
\label{app:2dkernel}
\begin{figure}[t]
	\centering
	\includegraphics[width=0.95\linewidth,height=0.65\textheight,keepaspectratio]{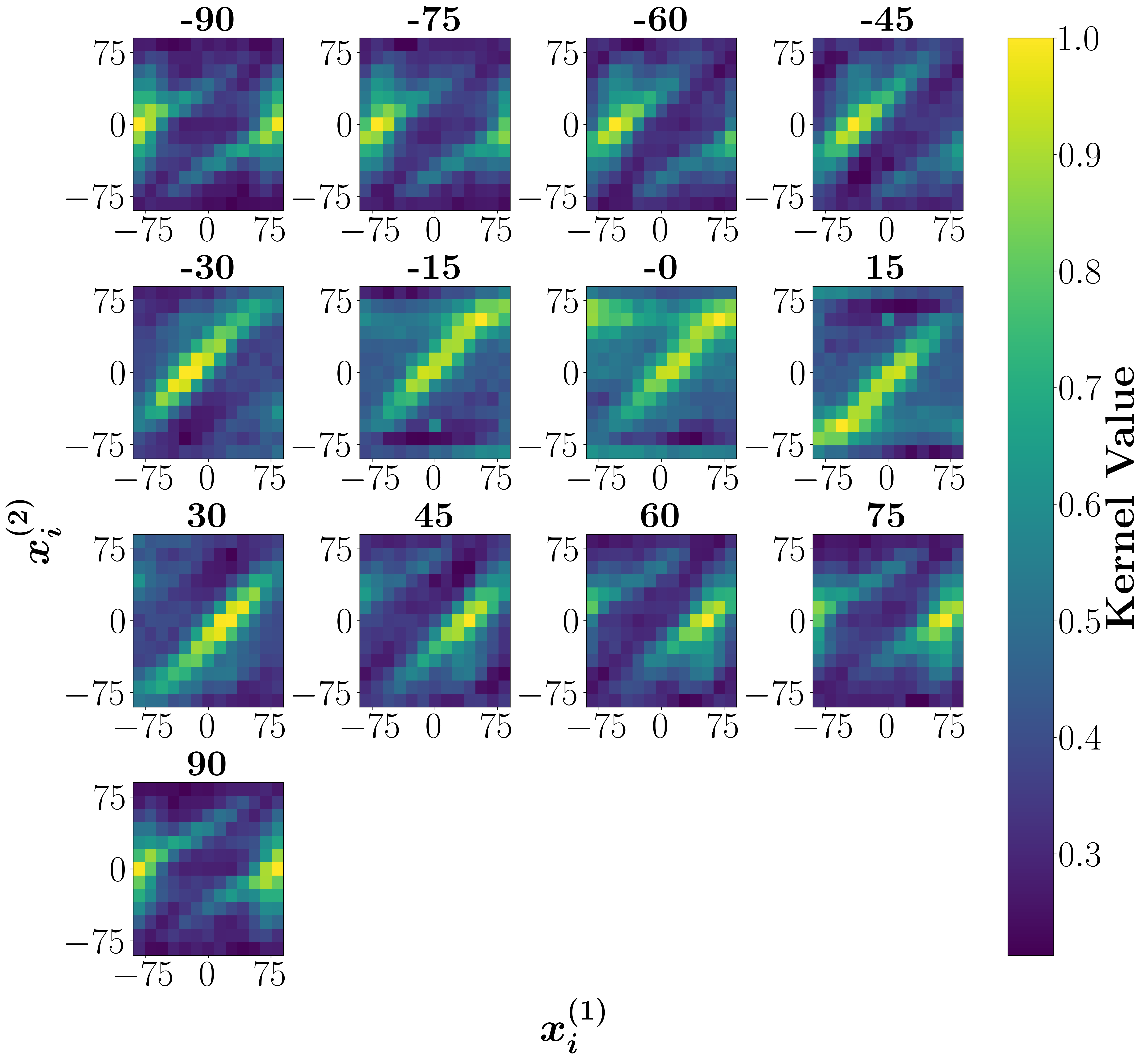}
	\caption{Experimental kernel slices for two-dimensional inputs. The title of each subplot indicates the value of $x_j^{(1)}$ in degrees. }
	\label{fig:2Dkernel}
\end{figure}

The kernel shown in Eq.~\eqref{eq:kernel_explicit} with the encoding unitary defined in Eq.~\eqref{eq:encoding_unitary} satisfies the symmetry \cite{kusumoto2021}  
\begin{multline}
	k\big(\{x_i^{(1)},x_i^{(2)}\},\{x_j^{(1)},x_j^{(2)}\}\big)\\
	= k\big(\{x_i^{(1)}-x_j^{(2)},x_i^{(2)}-x_j^{(2)}\},\{x_j^{(1)}-x_j^{(2)},0\}\big).
	\label{eq:2dkernel_symmetry}
\end{multline}
for two-dimensional inputs with the corresponding encoding unitaries as defined in \eqref{eq:encoding_unitary} in the main text. Given this, we can set $x_j^{(2)}=0$, and extract the kernel for two-dimensional inputs with only three independent parameters, $x_i^{(1)}$, $x_i^{(2)}$ and $x_j^{(1)}$. Fig. \ref{fig:2Dkernel} shows the $x_j^{(1)}$ slices of the kernel values obtained experimentally.

\section{Numerical and experimental kernel values for quantum data}
\label{app:qdata_kernel_values}
\begin{figure*}
    \centering
    \begin{minipage}[b]{0.49\linewidth}
        \centering
        \includegraphics[width=1\linewidth]{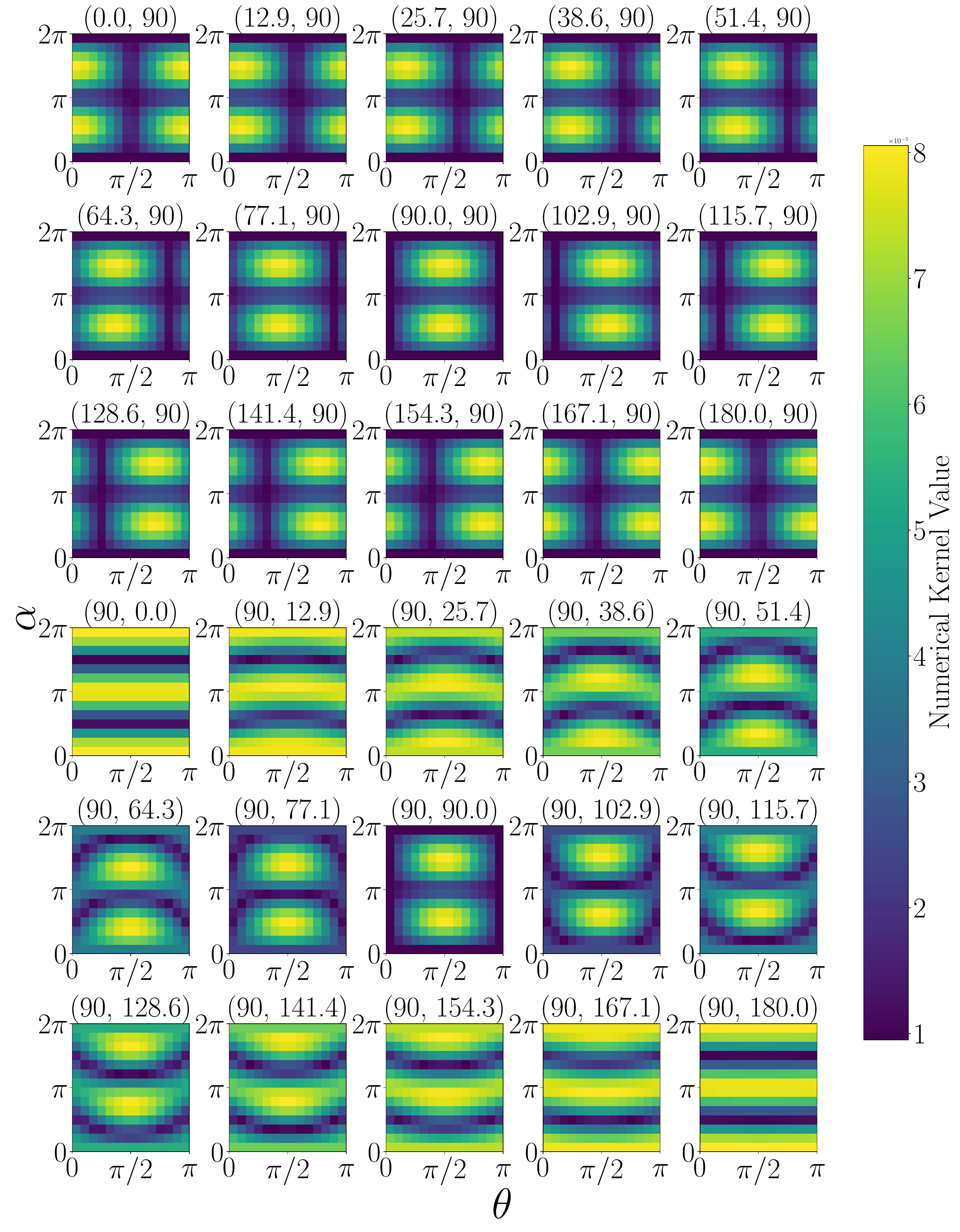}\\
        (a)
    \end{minipage}
    \begin{minipage}[b]{0.49\linewidth}
        \centering
        \includegraphics[width=\linewidth]{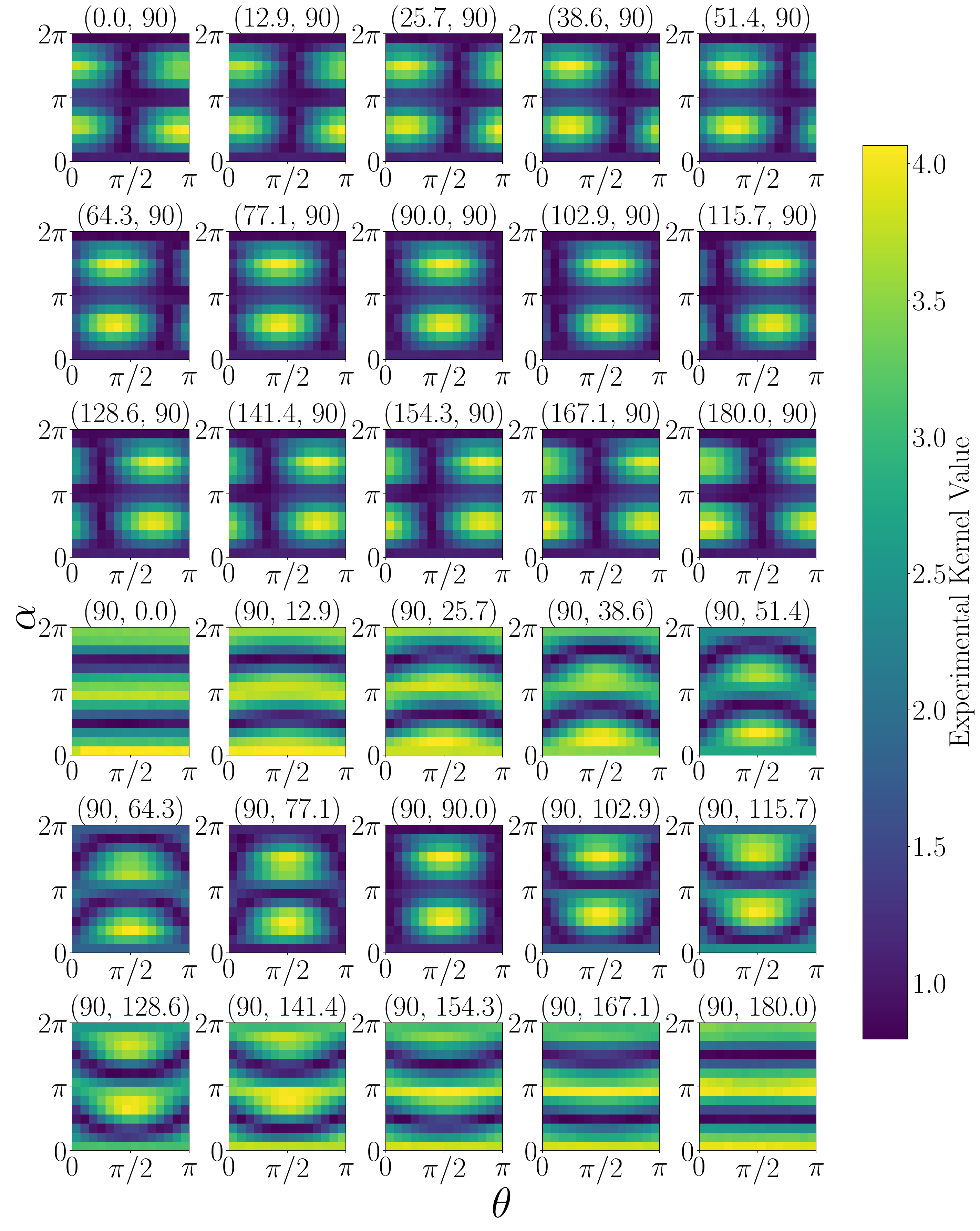}\\
        (b)
    \end{minipage}
    \caption{(a) The numerically simulated kernel values between all testing unitaries and each training unitary acting as quantum inputs. The title of each subplot indicates the $(\theta, \alpha)$ values of the training unitary {in degrees}. 
    (b) Experimental kernel values of all testing unitaries with respect to each training unitary. The title of each subplot indicates the $(\theta, \alpha)$ values of the training unitary in degrees.}
    \label{fig:kernel_values}
\end{figure*}

Numerically simulated kernel values between all testing unitaries and each training unitary acting as quantum inputs are shown in Fig. \ref{fig:kernel_values} (a). When compared to the experimentally obtained kernel values shown in Fig.~\ref{fig:kernel_values} (b), we see that the kernel values are identical in both cases up to an arbitrary scaling factor. This further validates our experimental setup and the results obtained. 

Fig. \ref{fig:kernel_values} (b) shows the kernel value of every test unitary relative to each training unitary (see Appendix~\ref{app:qdata_kernel_values} for the corresponding numerical results). Unitaries that induce ``similar transformations'' on the register --- {in particular, a similar capacity to generate entanglement} --- yield large kernel values, confirming that the kernel measures similarity directly in the operator space rather than in the raw parameter space \cite{markov2023quantumsimilarity}. 
Consider the training unitary $U(\pi/2,\pi/2)$ (subplot titled $(90,90)$): the corresponding heat map shows elevated kernel values not only near $(\pi/2,\pi/2)$ but also near $(\pi/2,3\pi/2)$, since $U(\pi/2,\pi/2)$ and $U(\pi/2,3\pi/2)$ induce the same transformation on the register up to an overall phase and are therefore indistinguishable to the kernel. This intrinsic invariance is precisely what allows the model to generalize beyond the training region of parameter space.

\section{Accuracy of quantum classification vs entangling time}
\label{app:qq_accuracy_vs_tau}
\begin{figure}
    \centering
    \begin{minipage}[b]{0.49\linewidth}
        \centering
        \includegraphics[width=1\linewidth]{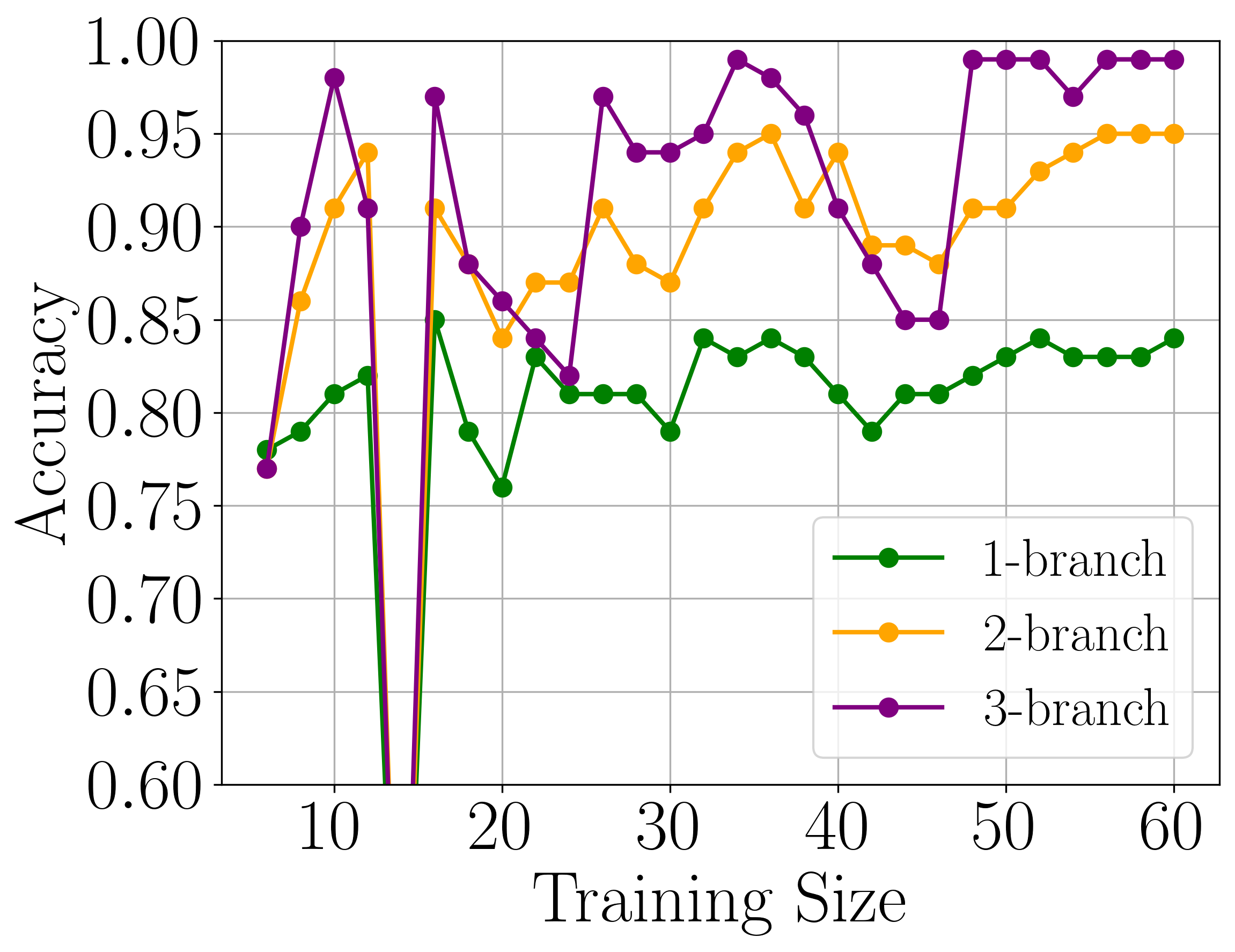}\\
        (a)
    \end{minipage}
    \begin{minipage}[b]{0.49\linewidth}
        \centering
        \includegraphics[width=1\linewidth]{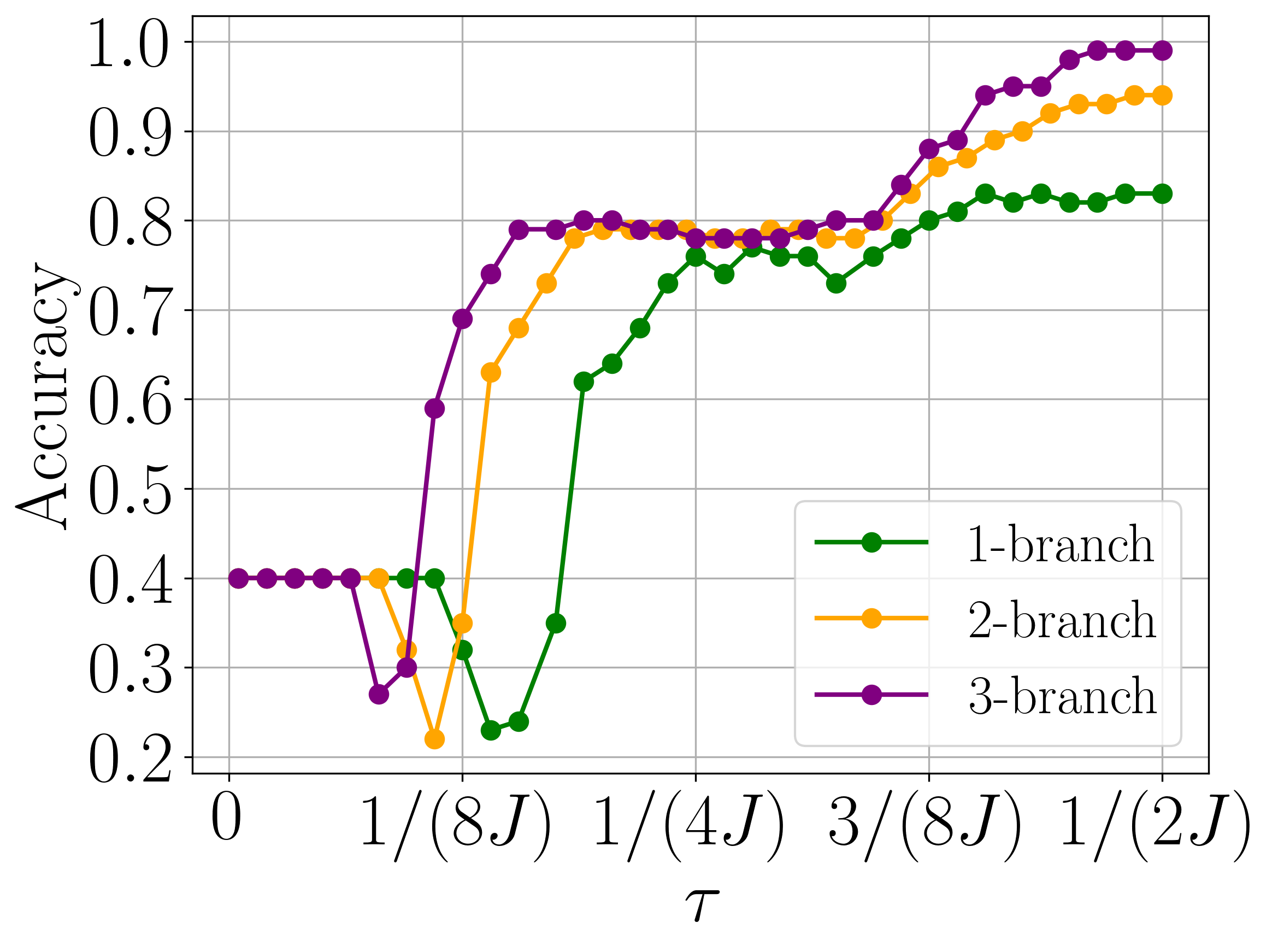}\\
        (b)
    \end{minipage}
    \caption{
    (a) The accuracy of the QKL SVM versus the number of training unitaries for 1-, 2-, and 3-branch star systems with a fixed training size of 34 unitaries. 
    (b) The accuracy versus the nonlocal evolution time $\tau$ used in the entangling unitary $U_e$ for 1-branch, 2-branch and 3-branch systems.}
    \label{fig:qq_accuracies}
\end{figure}

Fig. \ref{fig:qq_accuracies} (a) shows classification accuracy versus size of training-set for the 1-, 2-, and 3-branch systems, saturating at approximately 85\%, 95\%, and 99\%, respectively. This indicates that classification performance improves as the dimension of the feature space, set by the number of branches, increases. Fig. \ref{fig:qq_accuracies} (b) shows accuracy versus the evolution time $\tau$ in the nonlocal operation $\exp\left(-i \tau 2\pi J I_z^C I_z^A\right)$ in $U_e$.  Here, $J$ is the coupling between the C and A qubits.
The accuracy drops sharply as $\tau$ is decreased below $1/(2J)$, the time required for the CNOT-like gate to fully entangle C and A, underscoring the crucial role of $U_e$ in realizing the useful quantum kernel. 

\section{Hybrid workflow for QKL}
\label{app:workflow}
Here we describe the hybrid (quantum-classical) workflow that converts the raw NMR-measured kernel entries into the regression and classification plots shown in the main text.  All post-processing is carried out on a classical computer using Python\,3.11 with \texttt{NumPy}, \texttt{SciPy} and \texttt{scikit-learn} libraries.

\paragraph{Experimental kernel acquisition:}
For every pair of input data points $(\chi_i,\chi_j)$ we experimentally record the expectation value
$
k_{\mathrm{exp}}(\chi_i,\chi_j)=
\text{Tr}\!\bigl[
V^\dagger(\chi_j)V(\chi_i)\rho_0V^\dagger(\chi_i)V(\chi_j)\,I_z^{\mathrm C}
\bigr]
$.
This yields a symmetric, positive semidefinite matrix 
$
K\in\mathbb R^{N\times N}
$
(\(N\) is the number of training inputs). 

\paragraph{Normalization:}
We normalize the kernel matrix by the training-set maximum entry:
$K \leftarrow K / s$, where $s=\max_{i,j} K_{ij}$ computed on the training kernel matrix $K$, and the same $s$ is applied to test--train kernel values. 
We found that centering or trace/unit-diagonal normalizations degraded supervised performance in our setting with a smaller training size, whereas max-scaling preserved discriminative structure. 
Results were unchanged when replacing $\max$ by a robust percentile scale ({e.g.\ 99.5th percentile of elements of the kernel matrix}).

\paragraph{Regularization:}
The kernel matrix \(K\) is regularized to suppress numerical instabilities in the subsequent regression and classification tasks. The regularization is done with the help of the \texttt{scikit-learn} library. {In kernel ridge regression, the regularization parameter $\alpha$ sets the strength of the standard L2 (Tikhonov) regularization used to stabilize the fit and avoid overfitting. In SVM classification, the parameter $C$ is the soft-margin penalty parameter controlling the trade-off between maximizing the margin and allowing training misclassifications~\cite{scikit-learn}.} The optimal values of these parameters are determined by the cross-validation (CV) method for the classical tasks. However, for the quantum task, the training data are not randomly sampled or uniformly distributed in the domain of interest. The peculiar distribution of the training data does not allow us to find the best $C$-value through well-known CV methods. Hence, the SVM for the quantum task is not regularized beyond normalization. The parameters \(\alpha\) and $C$ in functions \texttt{KernelRidge} and \texttt{SVC} are set to the values specified below for each task.

\begin{itemize}
\item Regression (sine curve):  \(\alpha=10^{-4}\). 
\item Regression (seventh-degree polynomial): \(\alpha=0.05\)
\item Classification (circles \& moons):  \(C=0.55\) for circles and \(C=1.4\) for moons.
\end{itemize}

\paragraph{Prediction on dense grids:}
For visualization, the trained estimators require kernel values between \textit{test} inputs \(\tilde\chi\) and every
training point \(\chi_j\).  
Those values are \textit{not} measured on hardware; instead, they are \textit{interpolated} from the experimentally tabulated kernel signal.
The interpolation procedure maps a phase difference \(\phi=(\tilde\chi-\chi_j)/w\) onto the nearest tabulated value in the interpolated kernel data, {where \(w = \max_{i,j} |\chi_i-\chi_j| / (\tau/2)\), with $\tau$ being the period of $S(\varphi)$, which is $\pi$ in our case}. In this way, no additional calls to the spectrometer are required for plotting decision boundaries or regression curves. This step is not carried out for the quantum task since the data points are not classical.

\bibliographystyle{apsrev4-2}
\bibliography{references,sup_references}

\end{document}